\documentstyle[10pt]{article}
\makeatletter
\newdimen\normalarrayskip              
\newdimen\minarrayskip                 
\normalarrayskip\baselineskip
\minarrayskip\jot
\newif\ifold             \oldtrue            
\def\arraymode{\ifold\relax\else\displaystyle\fi} 
\def\eqnumphantom{\phantom{(\theequation)}}     
\def\@arrayskip{\ifold\baselineskip\z@\lineskip\z@
     \else
     \baselineskip\minarrayskip\lineskip2\minarrayskip\fi}
\def\@arrayclassz{\ifcase \@lastchclass \@acolampacol \or
\@ampacol \or \or \or \@addamp \or
   \@acolampacol \or \@firstampfalse \@acol \fi
\edef\@preamble{\@preamble
  \ifcase \@chnum
     \hfil$\relax\arraymode\@sharp$\hfil
     \or $\relax\arraymode\@sharp$\hfil
     \or \hfil$\relax\arraymode\@sharp$\fi}}
\def\@array[#1]#2{\setbox\@arstrutbox=\hbox{\vrule
     height\arraystretch \ht\strutbox
     depth\arraystretch \dp\strutbox
     width\z@}\@mkpream{#2}\edef\@preamble{\halign \noexpand\@halignto
\bgroup \tabskip\z@ \@arstrut \@preamble \tabskip\z@ \cr}%
\let\@startpbox\@@startpbox \let\@endpbox\@@endpbox
  \if #1t\vtop \else \if#1b\vbox \else \vcenter \fi\fi
  \bgroup \let\par\relax
  \let\@sharp##\let\protect\relax
  \@arrayskip\@preamble}
\def\eqnarray{\stepcounter{equation}%
              \let\@currentlabel=\theequation
              \global\@eqnswtrue
              \global\@eqcnt\z@
              \tabskip\@centering
              \let\\=\@eqncr
              $$%
 \halign to \displaywidth\bgroup
    \eqnumphantom\@eqnsel\hskip\@centering
    $\displaystyle \tabskip\z@ {##}$%
    &\global\@eqcnt\@ne \hskip 2\arraycolsep
         $\displaystyle\arraymode{##}$\hfil
    &\global\@eqcnt\tw@ \hskip 2\arraycolsep
         $\displaystyle\tabskip\z@{##}$\hfil
         \tabskip\@centering
    &{##}\tabskip\z@\cr}
\makeatother

\def\beq{\begin{equation}}
\def\eeq{\end{equation}}
\def\bea{\begin{eqnarray}}
\def\eea{\end{eqnarray}}

\def\stackreb#1#2{\mathrel{\mathop{#2}\limits_{#1}}}
\begin{document}

\begin{titlepage}
\begin{center}
{\it P.N.Lebedev Institute preprint} \hfill FIAN/TD-15/92\\
{\it I.E.Tamm Theory Department} \hfill hepth@xxx/92\#\#
\begin{flushright}{September 1992}\end{flushright}
\vspace{0.1in}{\Large\bf Topological versus Non--Topological Theories\\
and $p-q$ Duality in $c \le 1$ 2d Gravity Models}\\[.4in]
{\large  S. Kharchev, A. Marshakov
\footnote{talk given by A.M. at the  International Workshop on String Theory,
Quantum Gravity and the Unification of Fundamental Interactions,  Rome,  21 --
26 September 1992.}
}\\
\bigskip {\it  P.N.Lebedev Physics
Institute \\ Leninsky prospect, 53, Moscow, 117 924, Russia,
\footnote{E-mail address: tdparticle@glas.apc.org}
}

\end{center}
\bigskip \bigskip

\begin{abstract}
We discuss the non--perturbative formulation for $c \leq 1$ string theory.
The field theory like formulation of topological and non--topological models
is presented. The integral representation for arbitrary $(p,q)$ solutions is
derived which explicitly obeys $p-q$ duality of these theories. The exact
solutions to string equation and various examples are also discussed.
\end{abstract}

\end{titlepage}

\newpage
\setcounter{footnote}0

\section{Introduction}

Recent years brought us to a great progress in
understanding of non-perturbative string theory. The key idea, established at
least for the most simple set of  $c \leq  1$  conformal theories interacting
with two-dimensional gravity, is the appearance of the structure of integrable
hierarchy in the description of generating function for physical correlators
in these models \cite{Douglas,FKN1}.

Fortunately, the particular solutions to non-perturbative string theory can be
singled from the whole set of solutions to the Kadomtsev-Petviashvili (KP) or
rather Toda lattice hierarchy by an additional requirement usually known in
the form of
string equation, which allows to present these particular solutions in a
conventional ``field-theory-like" form. Indeed, it serious of papers
\cite{KMMMZ91a,KMMMZ91b,KMMM92a,KMMM92b,Mar92} it was shown, that there exists
a certain matrix model,
describing the particular subset of solutions to (reduced) KP-hierarchy
satisfying at the same time the string equation. The proposed matrix theory can
be considered as unifying theory for  $c \leq  1$  coupled to 2d gravity string
models, allowing one to interpolate among them
\cite{KMMMZ91a,KMMMZ91b,KMMM92b}, thus being a sort of effective string field
theory \cite{Mar92}.

Below, we are going to investigate solutions to various $(p,q)$-models (with
central charges  $c_{p,q}=1-6{(p-q)^2\over pq} )$  coupled to $2d$ gravity in
more details. Moreover, we would stress the advantages of matrix integral (or
better multiple integral
\footnote{Indeed, the naive very simple matrix form for such solutions might
not be
really their basic feature. For example, in the case of ordinary discrete
matrix models the generalization to ``multimatrix case" is better done via
{\it conformal multimatrix models} which do not have an elegant representation
by {\it matrix} integral, but are {\it multiple} integral solutions to the
extended (discrete) Virasoro-$W$ constraints \cite{KMMMP92}.
}
) representation for the particular ``stringy" solutions to KP hierarchy.

The basic feature of these matrix integrals is that they give an explicit
solution to string equation around a {\it topological} point. By definition,
for the simplest case of $(p=2,q=1)$ theory such integral was derived by
Kontsevich \cite{K1,K2} when studying topological characteristics of 2d
gravity module
space. In its basic form it obeys all the properties of topological theory --
trivial continuum limit (the size of matrix  $N \rightarrow  \infty )$ means
actually  $N$-independence while  $N$  can be interpreted as a cutoff
parameter, very simple form of particular solutions (Airy function etc), deep
interrelation with the Landau-Ginzburg topological theories and good
quasiclassical properties \cite{KMMM92b}. However, the problem of higher
critical points is much more complicated question.

Below, we are going to argue, that it also gets much better understanding
along this line. Shortly, higher critical points can be described using the
same ``action" principle, based on study of the quasiclassical limit
\cite{Kri,TakTak}, but the exact answer has much more complicated form.
It means that the
higher critical points are no more {\it topological} theories in the naive
sense we used above.
\footnote{Of course, they could still be reminisents of some more complicated
topological theories (and their module spases) like $W$-gravity etc (see for
example \cite{GLM} and J.-L.Gervais's contribution to this volume).}
The integral representation for these solutions besides
the ``action" functional has very complicated structure of integration measure.
Nevertheless, this integral representation obeys the basic property of $p-q$
duality in the spirit of \cite{FKN3} and might turn to be useful for studying
the exact solutions in various cases.

In sect.2 we are going to repeat the main results of \cite{KMMM92b} on
topological
solutions and speculate on naive ``analytic continuation" to higher critical
points. In sect.3 we will formulate the ``action" principle and derive an
integral representation for arbitrary $(p,q)$-solutions. In sect.4 we consider
$p-q$ symmetry and present several simple and useful examples. Sect.5 contains
several examples of $c<1$ exact $(p,q)$-solutions and sect.6 -- some comments
on what is supposed to be a particular example of $c=1$ situation. In sect.7
we give several concluding remarks.

\section{Topological $(p,1)$ models}

First, we remind that the partition
function is defined \cite{KMMMZ91a,KMMMZ91b} as a matrix integral
\beq\label{1}
Z^{(N)}[V|M] \equiv  C^{(N)}[V|M] e^{TrV(M)-TrMV'(M)}\int
DX\ e^{-TrV(X)+TrV'(M)X}
\eeq
over  $N\times N$  ``Hermitean" matrices, with the normalizing factor given
by Gaussian integral
\beq\label{2}
C^{(N)}[V|M]^{-1} \equiv  \int \hbox{  DY }\ e^{-TrV_2[M,Y]},
$$
$$
V_2 \equiv \lim _{\epsilon \rightarrow 0}
{1\over \epsilon ^2}Tr[V(M+\epsilon Y) - V(M) - \epsilon YV'(M)]
\eeq
and  $Z$  actually depends on  $M$  only through the invariant variables
\beq\label{3}
T_k = {1\over k} Tr\ M^{-k}\hbox{, }  k\geq 1\hbox{  ;}
\eeq
moreover, if rewritten in terms of $T_k$,  $Z[V|T] = Z^{(N)}[V|M]$  is
actually independent of the size  $N$  of the matrices.

As a function of  $T_k$  $Z[V|T_k]$ is a  $\tau $-function of KP-hierarchy,
$Z[V|T_k] = \tau _V[T_k]$, while the potential  $V$  specifies (up to certain
invariance) the relevant point of the infinite-dimensional Grassmannian.

For various choices of the potential  $V(X)$  the model (\ref{1}) formally
reproduces various
$(p,q)$-series:  the potential $V(X) = {X^{p+1}\over p+1}$  can be associated
with the entire set of $(p,q)$-minimal string models with all possible $q$'s.
In order to specify $q$ one needs to make a special choice of $T$-variables:
all  $T_k= 0$, except for  $T_1$ and  $T_{p+q}$ (the symmetry between $p$ and
$q$ is implicit in this formulation).

However, this is only a formal consideration. For the potential  $V(X) =
{X^{p+1}\over p+1}$  the partition function  $Z[V|T_k] = \tau _V[T_k] \equiv
\tau _p[T_k]$  satisfies the string equation which looks like

\beq\label{4}
\sum ^{p-1}_{k=1}k(p-k)T_kT_{p-k} + \sum ^\infty _{k=1}(p+k)(T_{p+k} -
{p\over p+1}\delta _{k,1}) {\partial \over \partial T_k} \log \ \tau _p[T] = 0
\eeq
$i.e$. $\tau $-function is defined with all Miwa times (\ref{3}) around zero
values
(in $1/M$ decomposition like in original Kontsevich model) with the only
exception - $T_{p+1}$ is shifted what corresponds obviously to $(p,1)$ model.
Thus, we see that the matrix integral gives an explicit solution to
$(p,1)$ string models which should have mentioned above topological properties
and must be nothing but particular topological matter coupled to topological
gravity.

Of course, we still have an opportunity for analytic continuation in string
equation, using the definition of Miwa's times (\ref{3}). We have to satisfy
the following conditions:
\beq\label{5}
T_1 = x
$$
$$
T_2 = 0
$$
$$
...
$$
$$
T_{p+1} - {p\over p+1} = 0
$$
$$
T_{p+q} = t_{p+q} = fixed
$$
$$
T_{p+q+1} = 0
$$
$$
...
\eeq
which is a system of equations on the Miwa parameters  $\{\mu _i\}$,  $i =
1,...,N$. So, to do this analytical continuation one has to decompose the
whole set
\beq\label{6}
\{\mu _i\} = \{\xi _a\} \oplus  \{\mu '_s\}
$$
$$
T_k = {1\over k} TrM^{-k} = {1\over k} \sum ^N_{j=1}\mu ^{-k}_j = {1\over k}
\sum    \xi ^{-k}_a + {1\over k} \sum ^{N'}_{j=1}{\mu '}_j^{-k} \equiv
T^{(cl)}_k + T'_k
\eeq
into ``classical" and ``quantum" parts respectively. In principle it is clear
that we have now to solve the equations
\beq\label{7}
T^{(cl)}_k = {1\over k} \sum    \xi ^{-k}_a = t_{p+q}\delta _{k,p+q} -
{p\over p+1}\delta _{k,p+1}
\eeq
and this can be done adjusting a certain block form of the matrix  $M$
\cite{KMMMZ91b,Mar92}. However, in such a way we can only vanish several first
times,
and the rest ones can be vanished only adjusting correct behaviour in the limit
$N \rightarrow  \infty $. The most elegant way to do this
\footnote{suggested by A.Zabrodin}
is to look at the formula
\beq\label{8}
\exp  (- \sum ^\infty _{k=1}\lambda ^kT^{(cl)}_k) =
\lim _{K \rightarrow \infty }
(1 - {1\over K}\sum ^\infty _{k=1}\lambda ^kT^{(cl)}_k)^K = \prod  _a (1 -
{\lambda \over \xi _a})
\eeq
and then the solution to (\ref{7}) will be given by  $K$  sets of roots of the
equation
\beq\label{9}
\sum ^\infty _{k=1}\lambda ^kT^{(cl)}_k - K = t_{p+q}\lambda ^{p+q} -
{p\over p+1}\lambda ^{p+1} - K = 0
\eeq
Obviously, the eigenvalues  $\xi _a$ will now depend on the size of the matrix
$N = (p+q)K + N'$  through explicit  $K$-dependence  $(\xi _a \sim
K^{1/(p+q)})$  and we lose one of the main features of $(p,1)$
theories mentioned above --- trivial dependence of the size of the matrix. Now
we can consider only matrices of {\it infinite} size and deal only with the
infinite determinant formulas.

That is why we call such way to get higher critical points as a formal one.
Below we will try to understand an alternative way of thinking, connected with
so-called $p$-times. Indeed, it was noticed in \cite{KMMM92b} that there exists
{\it a priori} another integrable structure in the model (\ref{1}), connected
with time
variables, related to the non-trivial coefficients of the potential  V. As a
results, the cases of monomial potential  $V_p(X) = {X^{p+1}\over p+1}$  and
arbitrary polynomial of the same degree  $(p+1)$  are closely connected with
each other. The direct calculation shows (see \cite{KMMM92b} for details)
\beq\label{10}
Z[V|T_k] = \tau _V[T_k] =
$$
$$
= \exp \left( - {1\over 2}\sum    A_{ij}(t)(\tilde T_i-t_i)(\tilde T_j-t_j)
\right)  \tau _p[\tilde T_k- t_k]\hbox{  ,}
\eeq
where
\beq\label{11}
V(x) = \sum _{i=0}^p {v_i\over i} x^i
$$
$$
\tilde T_k = {1\over k}Tr \tilde M^{-k}\hbox{ ,}
$$
$$
\tilde M^p = V'(M) \equiv  W(M)\hbox{  ,}
$$
$$
A_{ij} = Res_\mu  W^{i/p} dW^{j/p}_+\hbox{ ,}
\eeq
where $f(\mu )_+$ denotes the positive part of the Laurent series $f(\mu ) =
\Sigma \ f_i\mu ^i$ and
\beq\label{S.1}
\tau _p[T] \equiv  \tau _{V_p}[T]
\eeq
-- is the $\tau $-function of $p$-reduction. The parameters  $\{t_k\}$  are
certain linear combinations of the coefficients  $\{v_k\}$  of the potential
\cite{DVV,Kri}
\beq\label{12}
t_k = - {p\over k(p-k)}Res\ W^{1-k/p}(\mu )d\mu
\eeq
Formula (\ref{10}) means that ``shifted" by flows along $p$-times (\ref{12})
$\tau $-function is
easily expressed through the $\tau $-function of  $p$-reduction, depending only
on the difference of the time-variables $\tilde T_k$ and $t_k$. The change of
the spectral parameter in (\ref{5})  $M \rightarrow  \tilde M$  (and
corresponding
transformation of times  $T_k \rightarrow  \tilde T_k)$  is a natural step from
the point of view of equivalent hierarchies.

The $\tau $-functions in (\ref{10}) are defined by formulas
\beq\label{13}
\tau _V[T] = {\det \ \phi _i(\mu _j)\over \Delta (\mu )}
\eeq
and
\beq\label{14}
{\tau _p[\tilde T-t]\over \tau _p[t]} = {\det
\hat \phi _i(\tilde \mu _j)\over \Delta (\tilde \mu )}
\eeq
with the corresponding points of the Grassmannian determined by the basic
vectors
\beq\label{15}
\phi _i(\mu ) = [W'(\mu )]^{1/2} \exp \left( V(\mu ) - \mu W(\mu )\right)
\int   x^{i-1}e^{-V(x)+xW(\mu )} dx
\eeq
and
\beq\label{16}
\hat \phi _i(\tilde \mu ) = [p\tilde \mu ^{p-1}]^{1/2} \exp \left(
-\sum ^{p+1}_{j=1}t_j\tilde \mu _j\right)  \int
x^{i-1}e^{-V(x)+x\tilde \mu ^p}dx
\eeq
respectively. Then it is easy to show that $\hat \tau _p(T)$ satisfies the
$L_{-1}$- constraint with {\it shifted} KP-times in the following way
\beq\label{17}
\sum ^{p-1}_{k=1}k(p-k)(\tilde T_k-t_k)(\tilde T_{p-k}-t_{p-k}) +
\sum ^\infty _{k=1}(p+k)(\tilde T_{p+k}-t_{p+k}){\partial \over \partial %
\tilde T_k} \log  \hat \tau _p[\tilde T-t] = 0
\eeq
where $t_i$ defined by (\ref{12}) are {\it identically} equal to zero for  $i
\geq p+2.$

The formulas (\ref{10},\ref{17}) demonstrate at least two things. First, the
partition function
in the case of deformed monomial potential $(\equiv  polynomial$ of the same
degree) is expressed through the equivalent solution (in the sense
\cite{Shiota,Tak}) of the same $p$-reduced KP hierarchy, second -- not only
$t_{p+1}$ but all  $t_k$ with  $k \leq  p+1$  are not equal to zero in the
deformed situation. We will
call such theories as {\it topologically deformed $(p,1)$} models (in contrast
to {\it pure $(p,1)$} models given by monomial potentials  $V_p(X))$, the
deformation is ``topological" in the sense that it preserves all the features
of topological models we discussed above. Moreover, this ``topological"
deformation preserves almost all features of $2d$ Landau-Ginzburg theories
and from the point of view of continuum theory they should be identified
with the twisted Landau-Ginzburg topological matter interacting with gravity.

{}From the point of view of KP hierarchy we deal now again with  $p$-reduction.
Indeed, from eq.(\ref{14}) in the limit when all of the Miwa variables
$\tilde \mu _i$ go to infinity except of the first eigenvalue $(\tilde \mu _i
\equiv  \tilde \mu )$ one can obtain the expression for the Baker-Akhiezer
function which is almost equal to the first basic vector:
\beq\label{18}
\psi (\tilde \mu ,t) = \exp \left( \sum ^{p+1}_{j=1}t_j\tilde \mu _j\right)
\hat \phi _1(\tilde \mu ) = [p\tilde \mu ^{p-1}]^{1/2} \int
e^{-V(x)+x\tilde \mu ^p}dx
\eeq
where potential $V(x)$ is parameterized by $p$-times $\{t_k\}$ due to
eq.(\ref{12}).
It is evident that $\psi (\tilde \mu ,t)$ has the usual asymptotic
\beq\label{S.2}
\psi (\tilde \mu ,t) \stackreb{\mu \rightarrow \infty}{\rightarrow}
\exp \left( \sum ^{p+1}_{j=1}t_j\tilde \mu _j\right) \left( 1 +
O(\tilde \mu ^{-1})\right)
\eeq
Using equations of motion for {\it quasiclassical} KP hierarchy \cite{DVV,Kri}
\beq\label{S.3}
{\partial V\over \partial t_i} = - W^{i/p}_+
\eeq
(this is the consequence of parameterization (\ref{12})) one can easy to show
that
the Baker-Akhiezer function (\ref{18}) satisfies the usual equations of the
$p$-reduced KP hierarchy:
\beq\label{19}
[W({\partial \over \partial t_1}) + t_1]\psi (\tilde \mu ,t) =
\tilde \mu ^p\psi (\tilde \mu ,t)\hbox{  ,}
$$
$$
{\partial \psi \over \partial t_i} =
W^{i/p}_+({\partial \over \partial t_1})\psi \hbox{  .}
\eeq
where polynomials $W^{i/p}_+(\mu )$ are functions of $p$-times (\ref{12}). It
is important that $W^{i/p}_+(\mu )$ does not depend on $t_1$ for $i < p$ and,
therefore, in the corresponding Zakharov-Shabat equations we can treat
$\partial /\partial t_1$ as a formal {\it parameter}, not an operator. Thus,
we see that topologically deformed $(p,1)$ models which are quasiclassical
limit of the KP hierarchy in the sense of \cite{Kri} are simultaneously the
{\it exact}
solutions of the full $p$-KP hierarchy restricted on the ``small phase space"
\cite{DW}. The Baker-Akhiezer function (\ref{18}) represent the explicit
solution of
evolution equations along first $p$ flows and all basic vectors of the deformed
$(p,1)$ model can be obtained from $\psi (\tilde \mu ,t)$ with the help of the
formula
\beq\label{S.5}
\phi _i(\tilde \mu ,t) =
\exp \left( -\sum ^{p+1}_{j=1}t_j\tilde \mu _j\right)
{\partial ^{i-1}\psi (\tilde \mu ,t)\over \partial t^{i-1}_1}
\eeq
(Of course, the basic vectors of the pure $(p,1)$ model corresponding to
monomial potential $V_p$ can be obtained by setting $t_1 = ..$. $= t_p = 0$,
$t_{p+1} = {p\over p+1} )$. As a solution to string equation this deformed case
differs only in analytic continuation along first  $p$  times.

These topologically deformed $(p,1)$ models as we already said preserve all
topological properties of $(p,1)$ models. Indeed, according to \cite{FKN1}
shifting
of first times  $t_1,...,t_{p+1}$ is certainly not enough to get higher
critical points. To do this one has to obtain  $t_{p+q} \neq  0$, but this
cannot be done using above formulas naively, because it is easily seen from
definition (\ref{12}) of $p$-times, that  $t_k \equiv  0$  for  $k \geq  p+2$.
To do
this we have to modify the above procedure and we are going to this in next
section.

\section{Action principle}

The above scheme has a natural quasiclassical
interpretation. Indeed, the solution to $(p,1)$ theories given by the partition
function (\ref{1}) can be considered as a ``path integral" representation of
the solution to Douglas equations \cite{Douglas}
\beq\label{20}
[\hat P,\hat Q] = 1
\eeq
where  $\hat P$  and  $\hat Q$  are certain differential operators (of order
$p$  and  $q)$ respectively and obviously  $p-th$ order of  $\hat P$  dictates
$p$-reduction, while  $q$  stands for  $q-th$ critical point. Quasiclassically,
(\ref{20}) turns into Poisson brackets relation  \cite{Kri,TakTak}
\beq\label{21}
\{P,Q\} = 1
\eeq
where  $P(x)$  and  $Q(x)$  are now certain (polynomial) functions. It is
easily seen that the above case corresponds to the first order polynomial
$Q(x) \equiv  x$  and the $p$-th order polynomial  $P(x)$  should be identified
with  $W(x)\equiv V'(x)$  \cite{Kri}. Thus, the exponentials in
(\ref{1}), (\ref{15}) and (\ref{16}) acquire an obvious sense of action
functionals
\beq\label{22}
S_{p,1}(x,\mu ) = - V(x) + xW(\mu ) = - \int ^x_0dy\ W(y)Q'(y) + Q(x)W(\mu )
$$
$$
W(x) = V'(x) = x^p + \sum ^p_{k=1}v_kx^{k-1}
$$
$$
Q(x) = x
\eeq
and we claim that the generalization to arbitrary $(p,q)$ case must be
\beq\label{23}
S_{W,Q} =  - \int ^x_0dy\ W(y)Q'(y) + Q(x)W(\mu )
$$
$$
W(x) = V'(x) = x^p + \sum ^p_{k=1}v_kx^{k-1}
$$
$$
Q(x) = x^q + \sum ^q_{k=1}\bar v_kx^{k-1}
\eeq
Now the ``true" co-ordinate is $Q$, therefore the extreme condition of action
(\ref{23}) is still
\beq\label{24}
W(x) = W(\mu )
\eeq
having  $x = \mu $  as a solution, and for extreme value of the action one gets
\beq\label{25}
\left.S_{W,Q} \right |_{x=\mu } = \int ^\mu _0 dy\ W'(y)Q(y) =
$$
$$
= \sum ^{p+q}_{k=-\infty }t_k\tilde \mu ^k
\eeq
where  $\tilde \mu ^p = W(\mu )$  and
\beq\label{26}
t_k \equiv  t^{(W,Q)}_k = - {p\over k(p-k)}Res\ W^{1-k/p}dQ\hbox{  .}
\eeq
We should stress that the extreme value of the action (\ref{23}),
represented in the form (\ref{25}), determines the quasiclassical (or
dispersionless)
limit of the $p$-reduced KP hierarchy \cite{Kri,TakTak} with $p+q-1$
independent flows. We have seen that in the case of topologically deformed
$(p,1)$ models the quasiclassical hierarchy is exact in the strict sense:
topological solutions satisfy the full KP equations and the first basic vector
is just the Baker-Akhiezer function of our model (\ref{1}) restricted to the
small phase
space. Unfortunately, this is not the case for the general $(p,q)$ models: now
the quasiclassics is not exact and in order to find the basic vectors in the
explicit form one should solve the original problem and find the exact
solutions of the full KP hierarchy along first $p+q-1$ flows. Nevertheless, we
argue that the presence of the ``quasiclassical component" in the whole
integrable structure of the given models is of importance and it can give, in
principle, some useful information, for example, we can make a conjecture that
the coefficients of the basic vectors are determined by the derivatives of the
corresponding {\it quasiclassical $\tau $}-function
\footnote{The difference between generic $(p,q)$ and $(p,1)$ cases is also
crucial from the point of view of topological nature. We can see here again
a distinction between what we call topological and naively non-topological
models. The complications in general $(p,q)$ case might be connected with
the fact that we use not the most convenient representation for these theories
(see also footnote on the second page)}.

Returning to eq.(\ref{26}) we immediately see, that now only for  $k \geq
p+q+1$ \ \ $p$-times are identically zero, while
\beq\label{27}
t_{p+q} \equiv  t^{(W,Q)}_{p+q} = {p\over p+q}
\eeq
and we should get a correct critical point adjusting all  $\{t_k\}$  with
$k < p+q$  to be zero. The exact formula for the Grassmannian basis vectors in
general case acquires the form

\beq\label{28}
\phi _i(\mu ) = [W'(\mu )]^{1/2} \exp ( - \left.S_{W,Q}\right |
_{x=\mu })  \int   d{\cal M}_Q(x)f_i(x) \exp \ S_{W,Q}(x,\mu )
\eeq
where  $d{\cal M}_Q(x)$  is the integration measure. We are going to explain,
that the integration measure for generic theory determined by two arbitrary
polynomials $W$  and  $Q$  has the form
\beq\label{29}
d{\cal M}_Q(z) = [Q'(z)]^{1/2}dz
\eeq
by checking the string equation. For the choice (\ref{29}) to insure the
correct asymptotics of basis vectors  $\phi _i(\mu )$  we have to take
$f_i(x)$  being functions (not necessarily polynomials) with the asymptotics
\beq\label{30}
f_i(x) \sim  x^{i-1}(1 + O(1/x))
\eeq

\section{$p$-reduction and the Kac-Schwarz operator}

To satisfy the string
equation, one has to fulfill two requirements: the reduction condition
\beq\label{31}
W(\mu )\phi _i(\mu ) = \sum  _j C_{ij}\phi _j(\mu )
\eeq
and the Kac-Schwarz \cite{KSch,Sch} operator action
\beq\label{32}
A^{(W,Q)}\phi _i(\mu ) = \sum    A_{ij}\phi _j(\mu )
\eeq
with
\beq\label{33}
A^{(W,Q)} \equiv  N^{(W,Q)}(\mu ){1\over W'(\mu )}
{\partial \over \partial \mu } [N^{(W,Q)}(\mu )]^{-1} =
$$
$$
= {1\over W'(\mu )} {\partial \over \partial \mu } - {1\over 2}
{W''(\mu )\over W'(\mu )^2} + Q(\mu )
$$
$$
N^{(W,Q)}(\mu ) \equiv  [W'(\mu )]^{1/2} \exp ( - \left.S_{W,Q}
\right |_{x=\mu })
\eeq
These two requirements are enough to prove string equation (see
\cite{KMMMZ91b} for details). The structure of action immediately gives us
that
\beq\label{34}
A^{(W,Q)}\phi _i(\mu ) = N^{(W,Q)}(\mu )\int   d{\cal M}_Q(z) Q(z)f_i(z)
\exp \ S_{W,Q}(z,\mu )
\eeq
and the condition (\ref{32}) can be reformulated as a $Q$-reduction property
of basis
$\{f_i(z)\}$
\beq\label{35}
Q(z)f_i(z) = \sum    A_{ij}f_i(z)
\eeq

Let us check now the reduction condition. Multiplying  $\phi _i(\mu )$  by
$W(\mu )$  and integrating by parts we obtain
\beq\label{35a}
W(\mu )\phi _i(\mu ) =
$$
$$
= N^{(W,Q)}(\mu )\int   d{\cal M}_Q(z) f_i(z)
{1\over Q'(z)} {\partial \over \partial z} [\exp \ Q(z)W(\mu )] \exp [-
\int ^z_0dy\ W(y)Q'(y)] =
$$
$$
= - N^{(W,Q)}(\mu )\int   d{\cal M}_Q(z) \exp [S_{W,Q}(z,\mu )] \left(
{1\over Q'(z)} {\partial \over \partial z} - {1\over 2} {Q''(z)\over Q'(z)^2} -
W(z) \right) f_i(z) \equiv
$$
$$
\equiv  - N^{(W,Q)}(\mu )\int   d{\cal M}_Q(z) \exp [S_{W,Q}(z,\mu )]
A^{(Q,W)}f_i(z)
\eeq
Therefore, in the ``dual" basis  $\{f_i(z)\}$  the condition (31) turns to be

\beq\label{35b}
A^{(Q,W)}f_i(z) = - \sum    C_{ij}f_j(z)
\eeq
with  $A^{(Q,W)} (\neq  A^{(W,Q)})$  being the ``dual" Kac-Schwarz operator
\beq\label{36}
A^{(Q,W)} = {1\over Q'(z)} {\partial \over \partial z} - {1\over 2}
{Q''(z)\over Q'(z)^2} - W(z)
\eeq

The representation (\ref{28}), (\ref{29}) is an exact integral formula for
basis vectors
solving the $(p,q)$ string model. It has manifest property of $p-q$ duality (in
general $W-Q$), turning the $(p,q)$-string equation into the equivalent
$(q,p)$-string equation.

Now let us transform (\ref{28}), (\ref{29}) into a little bit more explicit
$p-q$ form. As
before for $(p,1)$ models we have to make substitutions, leading to
equivalent KP solutions:
\beq\label{37}
\tilde \mu ^p = W(\mu )\hbox{, }    \tilde z^q = Q(z)
\eeq
Then we can rewrite (\ref{28}) as
\beq\label{38}
\hat \phi _i(\tilde \mu ) = [p\tilde \mu ^{p-1}]^{1/2} \exp \left( -
\sum ^{p+q}_{k=1}t_k\tilde \mu ^k\right)  \int   d\tilde z
[q\tilde z^{q-1}]^{1/2} \hat f_i(\tilde z) \exp \ S_{W,Q}(\tilde z,\tilde \mu )
\eeq
where action is given now by
\beq\label{39}
S_{W,Q}(\tilde z,\tilde \mu ) =  - \left [ \int ^{\tilde z}_0d\tilde y
q\tilde y^{q-1}W(y(\tilde y)) \right ]_+  + \tilde z^q \tilde \mu ^p
$$
$$
= \sum_{k=1}^{p+q} \bar t_k \tilde z^k + \tilde z^q \tilde \mu ^p
\eeq
In new coordinates the reduction conditions are
\beq\label{40}
\tilde \mu ^p\hat \phi _i(\tilde \mu ) = \sum  _j
\tilde C_{ij}\hat \phi _j(\tilde \mu )
$$
$$
\tilde z^q\hat f_i(\tilde z) = \sum  _j \tilde A_{ij}\hat f_j(\tilde z)
\eeq
and for the Kac-Schwarz operators one gets conventional formulas
\cite{KSch,Sch,KMMMZ91b}
\beq\label{41}
\tilde A^{(p,q)} = {1\over p\tilde \mu ^{p-1}}
{\partial \over \partial \tilde \mu } - {p-1\over 2p} {1\over
{\tilde \mu ^p}}
 + {1\over p} \sum ^{p+q}_{k=1}kt_k\tilde \mu ^{k-p}
$$
$$
\tilde A^{(q,p)} = {1\over q\tilde z^{q-1}} {\partial \over \partial \tilde z}
- {q-1\over 2q} {1\over {\tilde z^
q}} + {1\over q} \sum ^{p+q}_{k=1}k\bar t_k\tilde z^{k-q}
\eeq
where for $(q,p)$ models we have introduced the ``dual" times:
\beq\label{S.6}
\bar t_k \equiv  t^{(Q,W)}_k = {q\over k(q-k)}Res\ Q^{1-k/q}dW
\eeq
in particularly, $\bar t_{p+q} = - {q\over p} t_{p+q} = - {q\over p+q}$ . Now
string equations give correspondingly
\beq\label{42}
\tilde A^{(p,q)}\hat \phi _i(\tilde \mu ) = \sum
\tilde A_{ij}\hat \phi _j(\tilde \mu )
$$
$$
\tilde A^{(q,p)}\hat f_i(\tilde z) = -\sum
\tilde C_{ij}\hat f_j(\tilde z)
\eeq

By these formulas we get a manifestation of $p-q$ duality if solutions to $2d$
gravity.

\section{Examples: topological and non-topological theories}

Now, let us
consider briefly several explicit examples. First, for monomials  $W(x) = x^p$
and  $Q(x) = x^q$, $\tilde \mu  \equiv  \mu $,  $\tilde z \equiv  z$,
$\hat \phi _i \equiv  \phi _i$ and  $\hat f_i \equiv  f_i$, thus, the formulas
of the previous section will be
\beq\label{43}
\phi _i(\mu ) =  [p\mu ^{p-1}]^{1/2} \exp \left( - {p\over p+q}
\mu ^{p+q}\right) \times
$$
$$
\times  \int   dz [qz^{q-1}]^{1/2} f_i(z) \exp  \left( - {q\over p+q} z^{p+q} +
z^q\mu ^p\right)
\eeq
and the Kac-Schwarz operators acquire the most simple form
\beq\label{44}
A^{(p,q)} = {1\over p\mu ^{p-1}} {\partial \over \partial \mu } - {p-1\over 2p}
{1\over \mu ^p} + \mu ^q
$$
$$
A^{(q,p)} = {1\over qz^{q-1}} {\partial \over \partial z} - {q-1\over 2q}
{1\over z^q} - z^p
\eeq
For any $(p,q)$ theory with  $q>p$  the formula (\ref{43}) maps it onto the
corresponding ``dual" theory with  $q<p$  and vice versa.

In such way one can easily consider the $(p,1)$ topological theories as dual to
the higher critical points of the $(1,p)$ theory with the potential  $V_2(x) =
{1\over 2}x^2$, $W_2 = x$. For this theory the ``topological" solution is
trivial
(for example, the partition function is given by a Gaussian integral and equals
to unity) so the basis vectors are
\beq\label{45}
f^{(1,p)}_i(z) = z^{i-1}
\eeq
and the Kac-Schwarz operator
\beq\label{46}
A^{(1,p)} = {\partial \over \partial z} - z^p
\eeq
preserves reduction of the corresponding $(p,1)$ model in a trivial way
\beq\label{47}
A^{(1,p)}f^{(1,p)}_i(z) = [ {\partial \over \partial z} - z^p] z^{i-1} =
$$
$$
= - z^{i+p-1} + (i-1)z^{i-2} = - f^{(1,p)}_{i+p-1}(z) + (i-1)f^{(1,p)}_{i-1}(z)
\eeq
In this particular case we see how the duality formula turns the problem of
finding nontrivial basis of \cite{KSch,KMMMZ91a,KMMMZ91b,K2} to the trivial
basis in the Grassmannian (\ref{45}), corresponding to sphere.

In general case, we have no more the situation when a non-trivial problem
reduces to a trivial one. Moreover, it can be easily proven that for a generic
$(p,q)$ model the string equation reduces to a sort of higher hypergeometrical
equation giving rise to (linear combinations of) generalized hypergeometric
functions \cite{Bailey,Bateman} (for integral formulas and connection to
free-field representation see also \cite{MorVin} and references therein).

Indeed, we can obtain some particular solutions of the conditions (\ref{32})
concerning only the shift $T_{p+q} \rightarrow  T_{p+q} - t_{p+q}$ as follows.
Let us consider the $(p,q)$ model with $q = pn + \alpha $, $\alpha  = 1$, ...,
$p-1$; $n = 1, 2, ...$ Using condition of $p$-reduction we can choose the whole
basis in the form
\beq\label{48}
\phi _{i+pk} = \mu ^{pk}\phi _i\hbox{ , } i = 1\hbox{, ... , } p
\eeq
and therefore eq.(\ref{32}) give the system of equations for first $p$ vectors:
\beq\label{49}
A^{(p,q)}\phi _i = \phi _{i+pn+\alpha } = \mu ^{pn}\phi _{i+\alpha }\hbox{ , }
i = 1\hbox{, ... , } p
\eeq
where in the case under consideration
\beq\label{50}
A^{(p,q)} = {1\over p\mu ^{p-1}} {\partial \over \partial \mu } - {p-1\over 2p}
{1\over \mu ^p} + \mu ^{pn+\alpha } \equiv
$$
$$
\equiv  N^{(p,q)}(\mu ) {1\over p\mu ^{p-1}} {\partial \over \partial \mu }
[N^{(p,q)}(\mu )]^{-1}
\eeq
and
\beq\label{50a}
N^{(p,q)}(\mu ) = [p\mu ^{p-1}]^{1/2} \exp \left( - {p\over p+q}
\mu ^{p+q}\right)
$$
$$
q = pn + \alpha
\eeq
After the substitution
\beq\label{51}
\phi _i = \mu ^{i-1} N^{(p,q)}(\mu ) u_i(\mu )
\eeq
and changing the spectral parameter
\beq\label{52}
z = {p\over p+q} \mu ^{p+q}
\eeq
the system (\ref{50}) acquires the remarkably simple form
\beq\label{53}
A_i u_i = u_{i+\alpha }\hbox{ , } i = 1\hbox{, ... , } p\hbox{; } u_{j+p}
\equiv  u_j
\eeq
with
\beq\label{54}
A_i = {d\over dz} + {i-1\over p+q} {1\over z}\hbox{  .}
\eeq
Equations (\ref{53}) can be easily solved, say, by series expansion
\beq\label{55}
u_i(z) = \sum ^\infty _{j=0}u_{ij}z^j
\eeq
giving
\beq\label{56}
u_{1j} = \prod ^{p-1}_{k=0}[j + ({1\over p+q} - 1)k]^{-1}u_{1,j-p}
\eeq
which determines function  $u_1(z)$  (up to  $p$  arbitrary constants) and
others can be obtained by action of (\ref{54}). Up to these constants a
particular
solution will be determined by the following formula
\beq\label{57}
u_{1,np} = \prod ^n_{l=0} \prod ^{p-1}_{k=0} (pl + ({1\over p+q} - 1)k
)^{-1}
\eeq
from which follows that the solutions can be expressed through generic
hypergeometric functions  $_rF_s(z)$. We are going to return to this problem in
a separate publication \cite{KMMM-long}.

As a concrete example of the relation to the Hypergeometric equation, we
present here the solution to $(2,2k-1)$ model
\beq\label{58}
u_1(\mu ) = exp( -{2 \over {2k+1}} \mu ^{2k+1}) _1F_1 ({1 \over 4k+2},
{1 \over 2k+1}; {4 \over 2k+1} \mu^{2k+1})
\eeq
For the particular case of $k=1$ this reduces to the well-known solution
of the Kontsevich model \cite{K1} being a linear combination of the Airy
functions \cite{KMMMZ91a,KMMMZ91b,K2}.

Now, let us only finish with a remark, that formulas (\ref{53}) are practically
equivalent for all $(p,q=np+\alpha )$ series with different  $n$  and
$\alpha $. They give a manifestation for a certain cyclic
${\bf Z}_p$-symmetry. The only difference in solution to different
$(p,q)$-string equations is the exact order established within the multiplets
of order  $p$. The exact sense of this symmetry deserves further investigation.

\section{Remarks on $c \to 1$ limit}

Let us now make some comments on $c=1$ situation. From
basic point of view we need in generic situation to get the most general
(unreduced) KP or Toda-lattice tau-function satisfying some (unreduced) string
equation. In a sense this is not a limiting case for $c<1$ situation but rather
a sort of ``direct sum" for all (p,q) models. This reflects that in conformal
theory coupled to $2d$ gravity there is no more a big difference between $c<1$
and $c=1$ situations - quite different before this coupling.

However, there are several particular cases when one can construct a sort of
direct $c \to 1$ limit and which should correspond to certain highly
``degenerate" $c=1$ theories. From the general point of view presented above
these are nothing but very specific cases of $(p,q)$ string equations, and
they could correspond only to a certain very reduced subsector of $c=1$ theory.

Indeed, it is easy to see, that for two special cases $p = \pm q$ the equations
(\ref{53}) can be simplified drastically, actually giving rise to a single
equation instead of a system of them. Of course, these two cases don't
correspond to minimal series where one needs $(p,q)$ being coprime numbers.
However, we still can fulfill both reduction and Kac-Schwarz condition and
these solutions to our equations using naively the formula for the central
charge, one mightidentify with $c=1$ for $p=q$ and $c=25$ for $p=-q$.

Now, the simplest ``topological" theories should be again with $q=1$. For
such case
``$c=1$" turns to be equivalent to a discrete matrix model \cite{KMMM92a}
while ``$c=25$" is exactly what one would expect from generalization of
topological Kontsevich-Penner approach \cite{Mar92,DMP} (see also
R.Dijkgraaf's contribution to this volume). Indeed, taking {\it non-polynomial}
finctions, like
\beq
W(x) = x^{-\beta }
$$
$$
Q(x) = x^{\beta }
\eeq
the action would acquire a logariphmic term
\beq
S_{-\beta ,\beta } = - \beta logx + {x^{\beta } \over \mu^{\beta }}
\eeq
while equations (\ref{53}) give rise just to rational solutions. It is very
easy to see that $\beta = 1$ immediately gives ``Kontsevich-Penner" result,
which rather corresponds to ``dual" to $c=1$ situation with matter central
charge being $c_{matter}=25$ with a highly non-unitary realization of
conformal matter.

On the other hand, $p=q=1$ solution is nothing but a trivial theory,
which however becomes a nontrivial discrete matrix model for unfrozen
zero-time. Moreover, these particilar $p = \pm q$ solutions become nontrivial
only if one considers the Toda-lattice picture with negative times being
involved into dynamics of the effective theory. On the contrary, we know
that $c<1$ $(p,q)$-solutions in a sense trivially depends on negative times
with the last ones playing the role of symmetry of string equation
\cite{KMMM92a}. It means, that we don't yet understand
enough the role of zero and negative times in the Toda-lattice formulation.

\section{Conclusion}

Let us make some conclusive remarks. We tried to present in
the paper the exact mechanism of transitions among different $(p,q)$ solutions
of non-perturbative $2d$ gravity in the framework of general scheme proposed in
papers \cite{KMMMZ91a,KMMMZ91b,KMMM92a,KMMM92b,Mar92}. We demonstrated that
a naive
analytic continuation in the space of Miwa parameters though correct formally
leads to certain practical difficulties in explicit description of higher
critical points even in trivial topological situation. Instead, we demonstrated
a concrete scheme, which allow one to shift ``classical" counterparts of the KP
times, determined by the coefficients of the potential and by the choice of
right variable.

The corresponding integral representation is a direct consequence of the action
principle and can be interpreted as a certain field theory integral with a
highly nontrivial measure. It obeys manifest $p-q$ symmetry which is evident
and restores equivalence in motion along naively two different $p$- and $q$-
directions. Moreover, the appearance of higher degrees of polynomials can be
obtained by transformation from the higher critical points of lower $p$ models.

There is still a lot of open questions. Even in a dual to topological $(p,1)$
series model there exists nontriviality after  $\alpha logX$  term (and
negative times terms) are added to the potential. For the  $p=1$  model this
gives rise to a separate interesting problem -- the discrete Hermitean matrix
model \cite{KMMM92a} and the question is about interpretation of such
generalizations of nontrivial theories.

The other question is  more deep understanding of generic $c = 1$ situation
(which is not reduced to particular ``degenerate" cases considered in sect.6)
and the role of negative times: symmetry between positive and negative times,
the ``dissappearing" of negative times in $c<1$ case etc.
Naively, the duality formula leads to a Fredholm
equation on basis vector, which can be solved by using the Hermit polynomials,
giving rise to the trivialized situation of a discrete 1-matrix model. It is
also
quite interesting to study the quasiclassical limit of general $(p,q)$
solutions
and to compare them with topological theories. This might shed light to the
underlying topological structure of naively non-topological theories.

All these problems deserves further investigation and we are going to return to
them elsewhere.

\bigskip
We are deeply indebted to R.Dijkgraaf, L.Girardello, A.Lossev, A.Mironov and
A.Zabrodin for illuminating discussions. A.M. is grateful to the organizers of
the International Workshop on String Theory, Quantum Gravity and the
Unification of Fundamental Interactions for warm hospitality in Rome.

\end{document}